\documentclass[aps,prl,twocolumn,numerical,superscriptaddress,nofootinbib,showpacs,showkeys,hyperref,longbibliography]{revtex4}

\usepackage{graphicx,color}
\usepackage{amsmath,amssymb,amsfonts}
\usepackage[colorlinks=true,plainpages=true]{hyperref}

\def \be {\begin{equation}}
\def \ee {\end{equation}}
\def \ee  {\end{equation}}
\def \bea {\begin{eqnarray}}
\def \eea {\end{eqnarray}}

\def \roots{\sqrt{s_{_{NN}}}}

\def \GeVc {\mbox{$\mathrm{GeV} / c$}}

\begin{document}
\title{
Investigation of  the elliptic flow fluctuations of the identified particles using the A Multi-Phase Transport model
}
\medskip
\author{Niseem~Magdy} 
\email{niseemm@gmail.com}
\affiliation{Department of Physics, University of Illinois at Chicago, Chicago, Illinois 60607, USA}

\author{Xu~Sun} 
\affiliation{Department of Physics, University of Illinois at Chicago, Chicago, Illinois 60607, USA}

\author{Zhenyu~Ye} 
\affiliation{Department of Physics, University of Illinois at Chicago, Chicago, Illinois 60607, USA}

\author{Olga~Evdokimov} 
\affiliation{Department of Physics, University of Illinois at Chicago, Chicago, Illinois 60607, USA}

\author{Roy~A.~Lacey} 
\affiliation{Department of Chemistry, State University of New York, Stony Brook, New York 11794, USA}


\begin{abstract}
A Multi-Phase Transport (AMPT) model is used to study the elliptic flow fluctuations of identified particles using participant and spectator event planes. The elliptic flow measured using the first order spectator event plane is expected to give the elliptic flow relative to the true reaction plane which suppresses the flow fluctuations. However, the elliptic flow measured using the second-order participant plane is expected to capture the elliptic flow fluctuations.
Our study shows that the first order spectator event plane could be used to study the elliptic flow fluctuations of the identified particles in the AMPT model.  
The elliptic flow fluctuations magnitude shows weak particle species dependence and transverse momentum dependence. Such observation will have important implications for understanding the source of the elliptic flow fluctuations.
\end{abstract}
\maketitle


Many studies of the ultra-relativistic heavy-ion collisions at the Relativistic Heavy Ion Collider and the Large Hadron Collider show that an exotic state of matter named Quark-Gluon Plasma (QGP) is created in these collisions. A large number of studies are focused on identifying the dynamical evolution and the transport properties of the QGP.

In heavy-ion collisions, the produced particle azimuthal anisotropy measurements have been used in various studies to show the viscous hydrodynamic response of the QGP to the initial energy density spatial distribution produced in the early stages of the collisions~\cite{Heinz:2001xi,Hirano:2005xf,Huovinen:2001cy,Hirano:2002ds,Romatschke:2007mq,Luzum:2011mm,Song:2010mg,Qian:2016fpi,Magdy:2017ohf,Magdy:2017kji,Schenke:2011tv,Teaney:2012ke,Gardim:2012yp,Lacey:2013eia}.  The azimuthal anisotropy of the particles emitted relative to the reaction plane $\Psi_{R}$ can be described by the Fourier expansion~\cite{Voloshin:1994mz,Poskanzer:1998yz} of the final-state azimuthal angle $\phi $ distribution,
\begin{eqnarray}
\label{eq:1-1}
\frac{dN}{d\phi}  &\propto&   1+ 2 \sum^{\infty}_{n=1}\textit{v}_{n} cos\left[  n (\phi - \Psi_{R})   \right]   ,
\end{eqnarray}
%

The first Fourier harmonic, $v_{1}$, is the directed flow; $v_{2}$ is called the elliptic flow, and $v_{3}$ is the triangular flow, etc. A wealth of information on the characteristics of the QGP has been gained via the anisotropic flow studies of directed and elliptic flow.~\cite{Magdy:2019ojv,Adam:2019woz,Magdy:2018itt}, higher-order flow harmonics $v_{n > 2}$~\cite{Adamczyk:2017ird,Magdy:2017kji,Adamczyk:2017hdl,Alver:2010gr, Chatrchyan:2013kba}, flow~fluctuations~\cite{Alver:2008zza,Alver:2010rt, Ollitrault:2009ie} and different flow harmonics correlations~\cite{STAR:2018fpo,Adamczyk:2017hdl,Qiu:2011iv, Adare:2011tg, Aad:2014fla, Aad:2015lwa}.

Hydrodynamic studies suggest that anisotropic flow stems from the evolution of the medium in the presence of initial-state anisotropies, determined by the eccentricities $\varepsilon_{n}$. The $v_{2}$ and $v_{3}$ flow harmonics are recognized to be  linearly correlated to $\varepsilon_{2}$ and $\varepsilon_{3}$, respectively~\cite{Song:2010mg, Niemi:2012aj,Gardim:2014tya, Fu:2015wba,Holopainen:2010gz,Qin:2010pf,Qiu:2011iv,Gale:2012rq,Liu:2018hjh}. Therefore for these flow harmonics,
\begin{eqnarray}
\label{eq:1-2}
v_{n} = \kappa_{n} \varepsilon_{n},
\end{eqnarray}
where $\kappa_{n}$ encodes knowledge about the medium properties such as the specific shear viscosity ($\eta/s$) of the QGP.  Accurate extraction of $\eta/s$ requires certain restrictions on the initial-state models employed in such extractions. Such constraints can be achieved via measurements of the flow harmonics and the event-by-event flow fluctuations~\cite{Borghini:2000sa}. Flow fluctuations could be arising from several sources: one of which has attracted considerable attention is the initial eccentricity fluctuations~\cite{Miller:2003kd,Manly:2005zy,Voloshin:2006gz}.
Recent~theoretical studies have begun to take into account initial conditions that include energy density fluctuations, initial flow~\cite{Gardim:2011qn,Gardim:2012yp,Gale:2012rq}, and the full shear stress tensor~\cite{Schenke:2019pmk} at $\mu_{B}=0$ and at $\mu_{B}$ $>$ 0~\cite{Werner:1993uh,Shen:2017bsr,Akamatsu:2018olk,Mohs:2019iee}.  Also, the partonic structure inside the nucleons has been considered in Reference~\cite{Steinheimer:2008hr}.

Recently,~Reference~\cite{Martinez:2019jbu} presented more realistic event-by-event fluctuating initial conditions, Initial~Conserved Charges in Nuclear Geometry (ICCING), of not only the initial energy density profile but also the initial conserved charges of baryon number (B), strangeness (S), and electric charge (Q) density distributions. This work pointed out that while baryon number and electric charge have almost the same geometries to the energy density profile, the initial strangeness distribution is considerably more eccentric. 
{\color{black}
Such an effect predicts that the elliptic flow fluctuations will be larger for the strange and multi-strange hadrons. This effect can be detected experimentally via studying the elliptic flow fluctuations of the identified hadrons. 
}

The ratio between four-particles elliptic flow, $v_2\{4\}$, and the two-particles elliptic flow, $v_2\{2\}$,  is~often used to estimate the strength of the elliptic flow fluctuations as a fraction of the measured flow harmonic strength~\cite{Giacalone:2017uqx,Alba:2017hhe}. 
However, important caveats to studying the elliptic flow fluctuations using ($v_2\{4\}/v_2\{2\}$)  for the identified hadrons are, first, the demand for high statistical power, and second, the multi-strange hadron identification process~\cite{Adamczyk:2015ukd}. Consequently, the ratio of $v_2\{4\}/v_2\{2\}$  is of limited experimental use for carrying out these investigations for the multi-strange hadrons.

 In this work, we  investigate an alternative validation scheme, which employs the use of the first-order spectator event plane ,$\Psi^{\rm SP}_{1}$, along with the second-order event plane $\Psi^{\rm EP}_{2}$ to study the elliptic flow fluctuations of the identified hadrons.  Here, the underlying notion is that $v^{\rm SP}_{2}$ (with respect to the spectator first-order event plane) will reduce the elliptic flow fluctuations due to the strong correlations between the $\Psi^{\rm SP}_{1}$  and the  true reaction plane. Therefore, the ratio $v^{\rm SP}_{2}/v^{\rm EP}_{2}$ is expected to reflect the elliptic flow fluctuations.  

For RHIC highest energy and using the STAR detector, we propose a similar investigation to be performed using the first-order spectator event plane from spectator neutrons, measured by the zero-degree calorimeters (ZDC)~\cite{Adler:2001fq} and the second-order event plane using the new installed Event-Plane-Detector (EPD)~\cite{Adams:2019fpo}. Consequently, we think that conducting a similar experimental study will reveal important information about the elliptic flow fluctuations and will shed light on the ICCING scenario suggested in Reference~\cite{Martinez:2019jbu}. 


\section{Method}\label{Sec:2}
The current study is conducted with simulated events for Au+Au collisions at \mbox{$\sqrt{s_{NN}}$ = 200~GeV}, collected using the AMPT~\cite{Lin:2004en} model with the string-melting mechanism and hadronic cascade on.  
{\color{black}
The~AMPT model, which has been widely employed to study relativistic heavy-ion \mbox{collisions~\cite{Lin:2004en,Ma:2016fve,Ma:2013gga,Ma:2013uqa,Bzdak:2014dia,Nie:2018xog,Bzdak:2014dia,Magdy:2020xqs}}, includes four main dynamical components: initial condition, parton cascade, hadronization, and hadronic rescatterings.
The initial conditions take into account soft string excitations and the phase space distributions of minijet partons, which are produced by the Heavy-Ion Jet Interaction Generator model (HIJING)~\cite{Wang:1991hta} in which the Glauber model with multiple nucleon scatterings are used to define the heavy-ion collisions initial state.

The partons scatterings are handled according to the Zhang's Parton Cascade (ZPC) model~\cite{Zhang:1997ej}, which contain only two-body elastic scatterings with a cross-section defined as:
\begin{equation}
\frac{d\sigma}{dt}=\frac{9\pi\alpha^{2}_{s}}{2}(1+\frac{\mu^{2}}{s})\frac{1}{(t-\mu^{2})^{2}},
\label{q1}
\end{equation}
where $\alpha_{s}$ = 0.47 is the strong coupling constant, $\mu$ is the screening mass and  $s$ and $t$ are the Mandelstam variables.
In the AMPT with the string-melting mechanism, the excited strings and minijet partons are melted into partons. The partons scatterings will lead to local energy density fluctuations, which are equivalent to the local transverse density of participant nucleons.

In the string-melting version and when partons stop interacting with each other, a quark coalescence model is used to couple partons into hadrons. Consequently, the partonic matter is then converted into hadronic matter and the hadronic interactions are given by the A Relativistic Transport (ART) model~\cite{Li:1995pra}, which incorporates both elastic and inelastic scatterings for baryon--baryon, baryon--meson, and meson--meson interactions.
}

In this work, the centrality intervals are defined by selecting the impact parameter distribution, then the AMPT events are analyzed using (i) the event plane method and (ii) the multi-particle cumulant technique~\cite{Bilandzic:2010jr,Bilandzic:2013kga,Jia:2017hbm,Gajdosova:2017fsc}. Using both methods, particle of interest (POI) comes from pseudorapidities $|\eta| < 1$, {\color{black}which matches the STAR experiment pseudorapidity acceptance}, and with  transverse momentum $0.1 < p_T < 4.0$~\GeVc. 

The second-order event plane ($\Psi^{\rm EP}_{2}$), is estimated from the azimuthal distribution of final-state particles. The elliptic flow that will be obtained using this method will then be corrected with the corresponding event plane resolution ($\rm Res$( $\Psi^{\rm EP}_{2}$))~\cite{Poskanzer:1998yz}. The $\Psi^{\rm EP}_{2}$  is reconstructed in a pseudorapidity range of $2.5 < |\eta| < 4.5$, {\color{black}which matches the STAR experiment EPD acceptance}, and \mbox{$0.1 < p_{T} < 2.0$~\GeVc}:
\begin{eqnarray}
 \Psi_{2}^{\rm EP} = \frac{1}{2} \tan^{-1}\left[  \frac{\sum \omega_{i} \sin(2\phi_{i})}{\sum \omega_{i} \cos(2\phi_{i})} \right] ,
\end{eqnarray}
where $\phi_i$ is the final-state azimuthal angle of particle $i$, and $\omega_i$ is its weight. {\color{black}The weight is chosen to be equal to $p_T$}.  Also, the first order spectator plane $\Psi_{1}^{\rm SP}$ is constructed using the AMPT spectator $x$ and $y$ position information. Using the spectator or the event planes we can give the elliptic flow as:
\begin{eqnarray}
 v^{\rm EP}_{2} &=&   \frac{\langle \cos\left( 2 (\phi_{i} -  \Psi_{2}^{\rm EP}  ) \right)  \rangle}{Res(\Psi_{2}^{\rm EP})},\\
 v^{\rm SP}_{2} &=&   \frac{\langle \cos\left( 2 (\phi_{i} -  \Psi_{1}^{\rm SP}  ) \right)  \rangle}{Res(\Psi_{1}^{\rm SP})},
\end{eqnarray}
where $\rm Res$( $\Psi^{\rm EP}_{2}$) and $\rm Res$( $\Psi^{\rm SP}_{1}$) represent the resolution of the event planes. {\color{black}The event planes resolution is calculated using the two-subevent method~\cite{Poskanzer:1998yz}.}

On the other hand, the standard (subevents) cumulant methods framework is discussed in References~\cite{Bilandzic:2010jr,Bilandzic:2013kga,Jia:2017hbm,Gajdosova:2017fsc}.
In the  standard cumulant method, the $n$-particle cumulants are constructed using particles from the $|\eta|<1.0$ acceptance. 
Thus the constructed two-  and four-particle correlations can be written as:
\begin{eqnarray}\label{eq:2-1}
\langle v^{2}_{n} \rangle &=&  \langle  \langle \cos (n (\varphi_{1} -  \varphi_{2} )) \rangle  \rangle,
\end{eqnarray}
\begin{eqnarray}\label{eq:2-2}
\langle v^{4}_{n} \rangle &=&   \langle \langle \cos ( n \varphi_{1} + n \varphi_{2} -  n \varphi_{3} -  n \varphi_{4}) \rangle \rangle ,
\end{eqnarray}
where, $\langle \langle \, \rangle \rangle$ represents the average over all particles in a single event, and then in average over all events,  $n$ is the harmonic number and $\varphi_{i}$ expresses the azimuthal angle of the $i^{\rm th}$ particle. Then the four-particle elliptic flow harmonic can be given as:
\begin{eqnarray}\label{eq:2-2}
v^{4}_{2}\{4\} &=&   2~\langle v^{2}_{2} \rangle - \langle v^{4}_{2} \rangle.
\end{eqnarray}

{\color{black}
In general, when the flow fluctuation $\sigma$ is smaller than the true reaction plan elliptic flow $\langle v_{2} \rangle$ one can write~\cite{Snellings:2011sz,Voloshin:2007pc}:
 \begin{eqnarray}\label{eq:x1}
 v^{\rm SP}_{2}    &=& \langle v_{2} \rangle   \\
 v^{\rm EP}_{2}    &=& \langle v_{2} \rangle  + 0.5 \dfrac{\sigma^{2}}{\langle v_{2} \rangle}.
\end{eqnarray}

Then the ratio $v^{\rm SP}_{2}/v^{\rm EP}_{2}$ can be used to estimate the strength of the elliptic flow fluctuations as a fraction of the measured flow harmonic (large value of $v^{\rm SP}_{2}/v^{\rm EP}_{2}$ indicates less fluctuations whereas a smaller value indicates large fluctuations),
 \begin{eqnarray}\label{eq:x2}
\dfrac{v^{\rm SP}_{2}}{v^{\rm EP}_{2} }  &=&  \dfrac{\langle v_{2} \rangle}{\langle v_{2} \rangle  + 0.5 \dfrac{\sigma^{2}}{\langle v_{2} \rangle}} = \dfrac{1}{1 + 0.5 \left(  \dfrac{\sigma}{\langle v_{2} \rangle} \right)^{2}  }
\end{eqnarray}
}

 The reliability of this elliptic flow fluctuations extraction will depend on the strength of the correlations between the spectator plane and the reaction plane.

\begin{figure}[t]

\centering{
\includegraphics[width=1.0 \linewidth,angle=0]{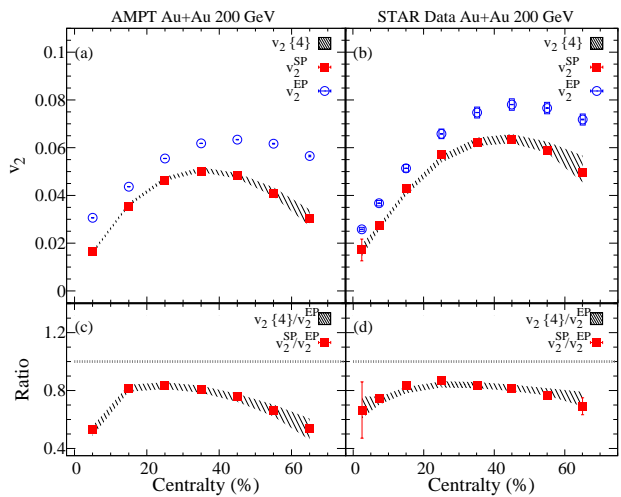}

\caption{
The charged particles centrality dependence of  $v^{\rm SP}_{2}$ and $v^{\rm EP}_{2}$ are compared to the four-particles elliptic flow (hashed band) for Au+Au collisions at $\roots$ = 200 GeV from the A Multi-Phase Transport (AMPT) model panel (\textbf{a}). The charged particles centrality dependence of  $v^{\rm SP}_{2}$ and $v^{\rm EP}_{2}$ are compared to $v_{2}\{4\}$ for Au+Au collisions at $\roots$ = 200 GeV from the STAR experiment~\cite{Wang:2005ab,Adam:2019woz} panel (\textbf{b}). The elliptic flow  fluctuations represented by the ratios $v^{\rm SP}_{2}/v^{\rm EP}_{2}$ and $v_{2}\{4\}/v^{\rm EP}_{2}$ are presented in panels (\textbf{c}, \textbf{d}).
\label{fig:fig1} 
 }
}
\end{figure}
\section{Results and Discussion}\label{Sec:3}

Panel (a) of Figure~\ref{fig:fig1} compares the centrality dependence of the four-particle elliptic flow ($v_{2}\{4\}$) with the elliptic flow measured with respect to the event plane ($v^{\rm EP}_{2}$) and spectators plane ($v^{\rm SP}_{2}$). The~comparison of the $v_{2}\{4\}$ and the $v^{\rm EP}_{2}$ shows larger $v^{\rm EP}_{2}$ magnitudes for  $v_{2}\{4\}$. By contrast, the~values for $v^{\rm SP}_{2}$ show good agreement with $v_{2}\{4\}$. Qualitatively, one expects such patterns due to the respective flow fluctuations contributions to $v_{2}\{4\}$ and $v^{\rm EP}_{2}$. The experimental measurements for charge hadrons reported by the STAR experiment, shown in Figure~\ref{fig:fig1}b~\cite{Wang:2005ab,Adam:2019woz}, also show good agreement between $v_{2}\{4\}$ and $v^{\rm SP}_{2}$($v^{\rm ZDC}_{2}$), consistent with the AMPT simulations. Here, no attempt was made to improve the agreement between the model and the experimental results by varying the model parameters to influence the flow magnitude and its associated fluctuations~\cite{Ma:2014xfa,Ma:2014pva,Singha:2016aim,Nayak:2017gpi}. We defer such an investigation to a future study.
The ratio $v^{\rm SP}_{2}/v^{\rm EP}_{2}$, presented in panel (c) from AMPT, and data panel (d) serves as a metric for elliptic flow fluctuations. The $v^{\rm SP}_{2}/v^{\rm EP}_{2}$  decrease from central to peripheral collisions, consistent with the patterns expected when initial-state eccentricity fluctuations dominate. Note,~however, that~other sources of fluctuations could contribute.

\begin{figure}[t]
\vskip -0.16cm
\centering{
\includegraphics[width=1.0 \linewidth, angle=0]{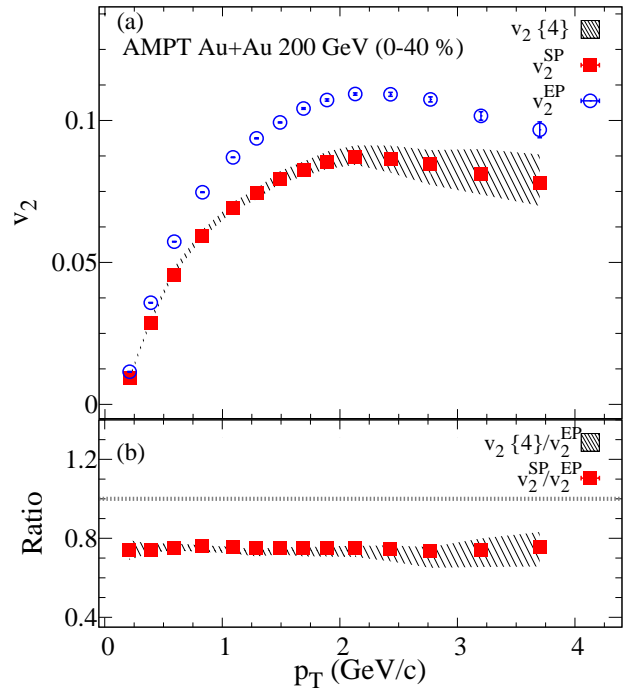}
\vskip -0.16cm
\caption{
The charged particles $p_{T}$ dependence of  $v^{\rm SP}_{2}$ and $v^{\rm EP}_{2}$ are compared to the four-particles elliptic flow (hashed band) panel (\textbf{a}).
The ratios $v^{\rm SP}_{2}/v^{\rm EP}_{2}$ and $v_{2}\{4\}/v^{\rm EP}_{2}$ are presented in panel (\textbf{b}) for Au+Au collisions at $\roots$ = 200 GeV from the AMPT~model.
\label{fig:fig2} 
 }
}
\end{figure}
\begin{figure}[!h]
\centering
\includegraphics[width=1.0 \linewidth, angle=0]{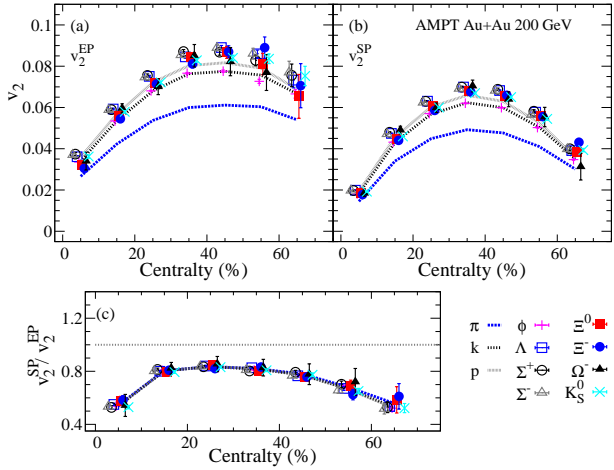}

\caption{
The identified particles centrality dependence of the elliptic flow harmonic with respect to participant and spectator event planes panels (\textbf{a},\textbf{b}) respectively. The elliptic flow  fluctuations represented by the ratio $v^{\rm SP}_{2}/v^{\rm EP}_{2}$ are presented in panel (\textbf{c}) for Au+Au collisions at $\roots$ = 200 GeV from the AMPT~model.
\label{fig:fig3}
 }
\vspace{-12pt}
\end{figure}

The transverse momentum dependence of the $v_{2}\{4\}$, $v^{\rm EP}_{2}$ and  $v^{\rm SP}_{2}$ are shown in Figure~\ref{fig:fig2}.  This~differential comparison further reflects the effect of the elliptic flow fluctuations on the $v^{\rm EP}_{2}$ which is highlighted in the ratio between $v^{\rm EP}_{2}$ and $v_{2}\{4\}$. Also a good agreement (within the errors) has been observed between the $v_{2}\{4\}$ and  $v^{\rm SP}_{2}$. The ratio $v^{\rm SP}_{2}/v^{\rm EP}_{2}$, presented in panel (b)  presents the strength of the elliptic flow fluctuations which shows no $p_{T}$ dependence, consistent with the preliminary STAR measurements~\cite{Niseem:QM2018}.

The centrality dependence of the identified particles $v^{EP}_{2}$ panel (a), $v^{SP}_{2}$ panel (b) and $v^{SP}_{2}/v^{EP}_{2}$ panel (c) are shown in Figure~\ref{fig:fig3} for Au+Au collisions at $\roots$ = 200 GeV from the AMPT model. 
The~results of $v^{EP}_{2}$ and $v^{SP}_{2}$ show the mass ordering effect on the observed magnitude. This mass ordering effect, which cancels out for the ratio $v^{SP}_{2}/v^{EP}_{2}$, presented in panel (c) indicates the domination of the initial-state eccentricity fluctuations in the AMPT model.

\begin{figure}[!h]
\centering{
\includegraphics[width=1.0 \linewidth, angle=0]{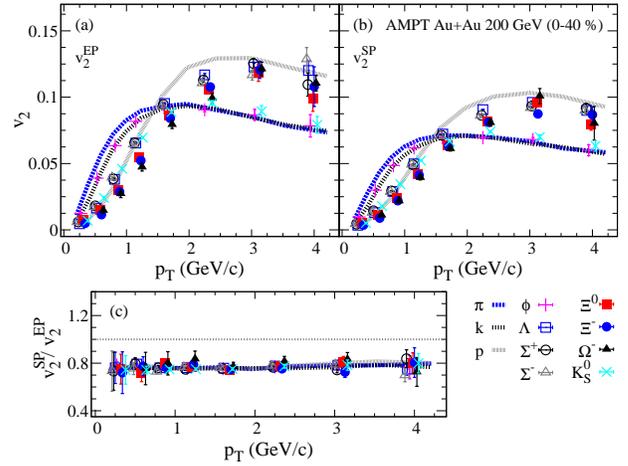}
\vskip -0.16cm
\caption{
 The identified particles $p_{T}$ dependence of the elliptic flow harmonic with respect to participant and spectator event planes panels (\textbf{a},\textbf{b}) respectively. The elliptic flow  fluctuations represented by the ratio $v^{\rm SP}_{2}/v^{\rm EP}_{2}$ are presented in panel (\textbf{c}) for Au+Au collisions at $\roots$ = 200 GeV from the AMPT~model.
\label{fig:fig4}
 }
}
\end{figure}

Figure~\ref{fig:fig4} compares the $p_{T}$ dependence of the identified particles $v^{EP}_{2}$ panel (a), $v^{SP}_{2}$ panel (b) and $v^{SP}_{2}/v^{EP}_{2}$ panel (c) for $0-40$\% Au+Au collisions at $\roots$ = 200 GeV from the AMPT model. The ratios $v^{SP}_{2}/v^{EP}_{2}$ panel (c) (elliptic flow fluctuations) show week sensitivity to the $p_{T}$ increase. The $v^{EP}_{2}$ and $v^{SP}_{2}$ vs.  $p_{T}$ show the expected mass ordering dependence, which cancels out for the ratio $v^{SP}_{2}/v^{EP}_{2}$ vs.  $p_{T}$, presented in panel (c), which further suggests that the elliptic flow fluctuations in the AMPT model are governed by initial-state fluctuations.
\vspace{-12pt}

\section{Conclusions}\label{Sec:4}
In summary, we studied the centrality and transverse momentum dependence of the identified particles $v^{SP}_{2}$, $v^{EP}_{2}$ and the elliptic flow fluctuations presented by the ratio $v^{SP}_{2}/v^{EP}_{2}$ using the AMPT model.
%
The magnitude of the elliptic flow fluctuations is observed to increase from central to mid-central collisions,  consistent with the patterns expected from the initial-state eccentricity fluctuations; a weak $p_{T}$  dependence is also observed.
The centrality and $p_{T}$ dependence of the identified particles $v^{EP}_{2}$ and $v^{SP}_{2}$ show the expected mass ordering. However, the elliptic flow fluctuations show no particle species dependence.
The integrated and differential elliptic flow fluctuation results indicate the domination of the effect of the initial-state eccentricity fluctuations as expected in the AMPT model. 
It is suggested that similar investigations of experimental data could display important insight on the ICCING scenario in heavy-ion collisions.

\section*{Acknowledgments}
The authors thank Jacquelyn Noronha-Hostler for the useful discussion and Emily Racow for the language check. This research was funded by the US Department of Energy under contract DE-FG02-94ER40865 (NM, XS, ZY and OE) and DE-FG02-87ER40331.A008 (RL).
%
\bibliography{ref} 
\end{document}